\documentclass[16ppt,twocolumn]{emulateapj}
\usepackage{amsmath,amssymb,rotfloat}
\usepackage{graphics}
\usepackage{graphicx}
\usepackage{color}
\usepackage{ulem}
\usepackage[section]{placeins}

\begin{document}

\shorttitle{MHD of colliding winds of WR-O stars}
\shortauthors{D. Falceta-Gon\c calves}
\title{Magnetic fields, non-thermal radiation and particle acceleration in colliding winds of WR-O stars}

\author{D. Falceta-Gon\c{c}alves\altaffilmark{1,2}}
\altaffiltext{1}{Escola de Artes, Ci\^encias e Humanidades, Universidade de S\~ao Paulo, Rua Arlindo Bettio, 1000, S\~ao Paulo, SP 03828-000, Brazil}
\altaffiltext{2}{SUPA, School of Physics \& Astronomy, University of St Andrews, North Haugh, St Andrews, Fife KY16 9SS, UK}

\begin{abstract}
Non-thermal emission has been detected in WR-stars for many years at long wavelengths spectral range, in general attributed to synchrotron emission. Two key ingredients are needed to explain such emissions, namely magnetic fields and relativistic particles. Particles can be accelerated to relativistic speeds by Fermi processes at strong shocks. Therefore, strong synchrotron emission is usually attributed to WR binarity. The magnetic field may also be amplified at shocks, however the actual picture of the magnetic field geometry, intensity, and its role on the acceleration of particles at WR binary systems is still unclear. In this work we discuss the recent developments in MHD modelling of wind-wind collision regions by means of numerical simulations, and the coupled particle acceleration processes related. 
\end{abstract}

\section{Introduction}

Wolf-Rayet (WR) stars represent a phase of the evolution of a massive star. 
Most of these objects are known to be in binary systems, making them a very 
interesting laboratory for the study of shocks in the space. This because 
the winds of massive stars are fast ($u_{\rm wind} \sim 1000 - 3000$km s$^{-1}$) 
and, in the case of luminous blue varibles (LBVs) and WR stars, very massive as 
well ($\dot{M} \sim 10^{-7} - 10^{-4}$M$_\odot$ yr$^{-1}$). When two (or more)
massive stars form an orbiting system their winds are likely to shock 
violently, resulting in a vast sort of interesting phenomena, e.g. 
strong x-ray emission and high energy physics \citep[see e.g.][]{usov93,pittard06,abraham10,reitberger14}. 

Among the most intriguing processes involving shocks at massive binary systems is 
particle acceleration. Synchrotron emission at large wavelengths have been observed 
in several WR binaries, revealing both the existence of relativistic particles and 
strong magnetic fields. Since current models of particle acceleration depends on the 
magnetization of the shocks, the study of the evolution of the magnetic fields in 
wind-wind collision regions is crucial.

In this work we present the main results of recent magneto-hydrodynamical numerical 
simulations of wind-wind collisions. We also briefly discuss the implications of such 
modelling on the understanding of particle acceleration in such objects.

\section{Modelling}

In order to obtain the dynamical evolution of the shock region of the massive binary system we 
have employed a number of magneto-hydrodynamical (MHD) numerical simulations. 
The simulations were performed solving the set of ideal MHD equations, in 
conservative form, as performed in \citet{falceta12}.
The set of equations is complete with an explicit 
equation for the radiative cooling. Two distinct setups were studied, namely the cases where 
radiative cooling were neglected and included, for comparison. These different setups 
correspond to the cases where the cooling time is longer, or shorter, respectively, to 
the dynamical timescales of the flows at the shock region. 

The initial setup used was chosen to represent the system of $\eta$ Carinae, with 
wind mass loss rates for the two stars as $\dot{M_1} \sim 2 \times 10^{-4}\; \rm 
{M}_{\odot}\; \rm {yr}^{-1}$ and $\dot{M_2} \simeq 10^{-5}\; \rm {M}_{\odot}\; \rm {yr}^{-1}$, 
and wind terminal velocity $v_1 \sim 700$ km s$^{-1}$ and $v_2 \simeq 3 \times 10^{3}$ km s$^{-1}$, 
respectively \citep{falceta05,abraham05a,abraham05b, falceta07, falceta08}. The orbit is set with a major 
semi-axis $a = 15$AU and eccentricity $e=0.9$. 
The magnetic field of both stars is initially set as a dipole with surface polar 
intensity $B_0 = 1$G, weak compared to the thermal and kinetic terms of the winds.

\section{Results}

The numerical integration of each model was performed for 2 complete orbital periods, 
which corresponds to approximately 11 years. Irrespectively to the model, the amplification 
of magnetic field at the shock has been shown to be larger than the values predicted by 
the Rankine-Hugoniot jump conditions \citep{silva15}. This occurs because the R-H jump 
conditions are computed for 1D solution of conservation equations, while the shock is 
multidimensional. It was already shown in \citet{falceta12} that the flow of the post-shocked 
gas occurs mostly parallel to the field lines, reducing the local pressure and allowing further 
compression of the magnetic fields at the shock layers. Also, in the cases of strong cooling, 
this compression occurs even further, resulting in magnetic fields orders of magnitude larger 
that the predicted by R-H jump conditions. 

\begin{figure}[H]
\begin{center}
\includegraphics[scale=0.2]{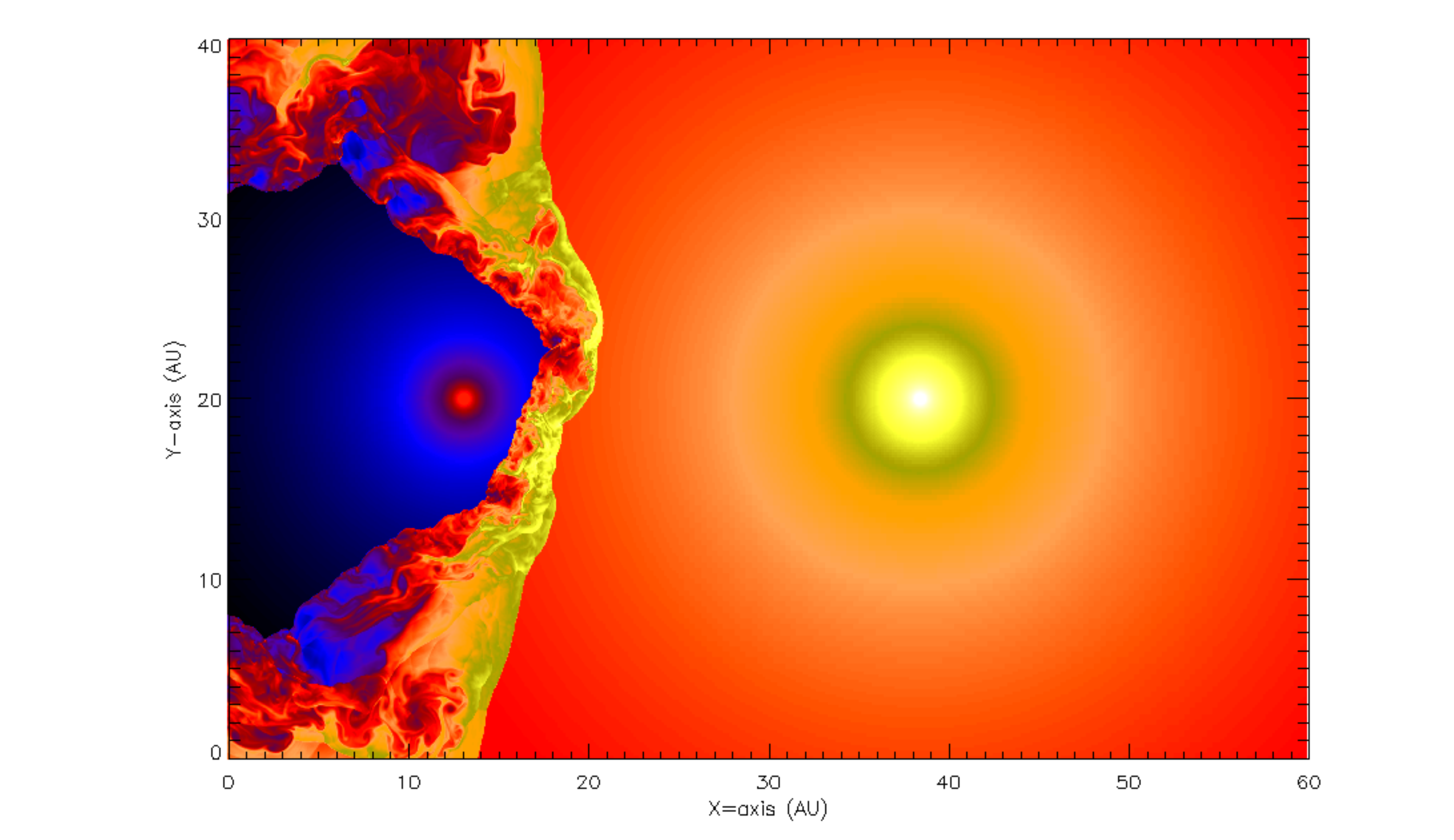} \\
\includegraphics[scale=0.2]{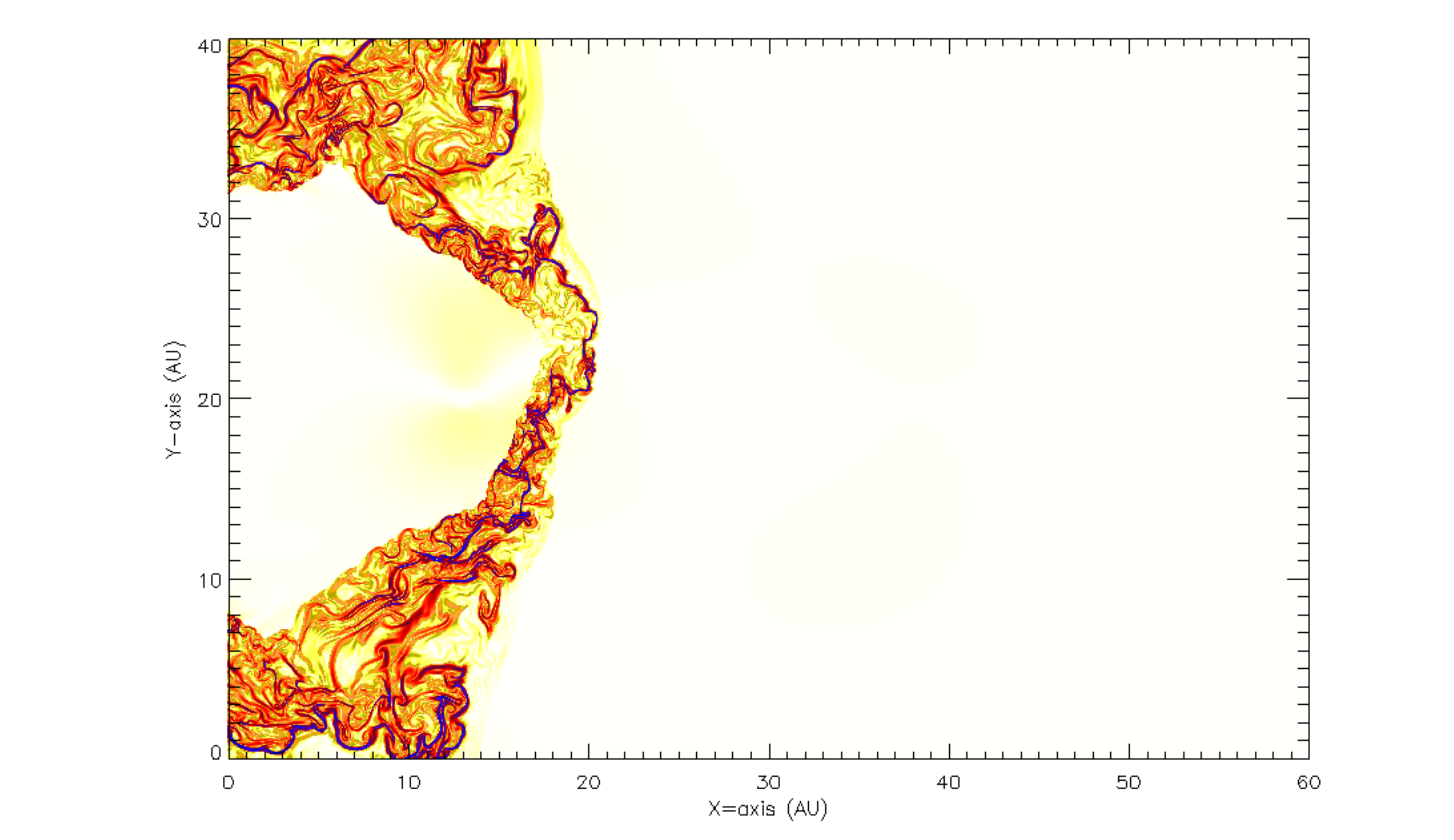} \\
\includegraphics[scale=0.2]{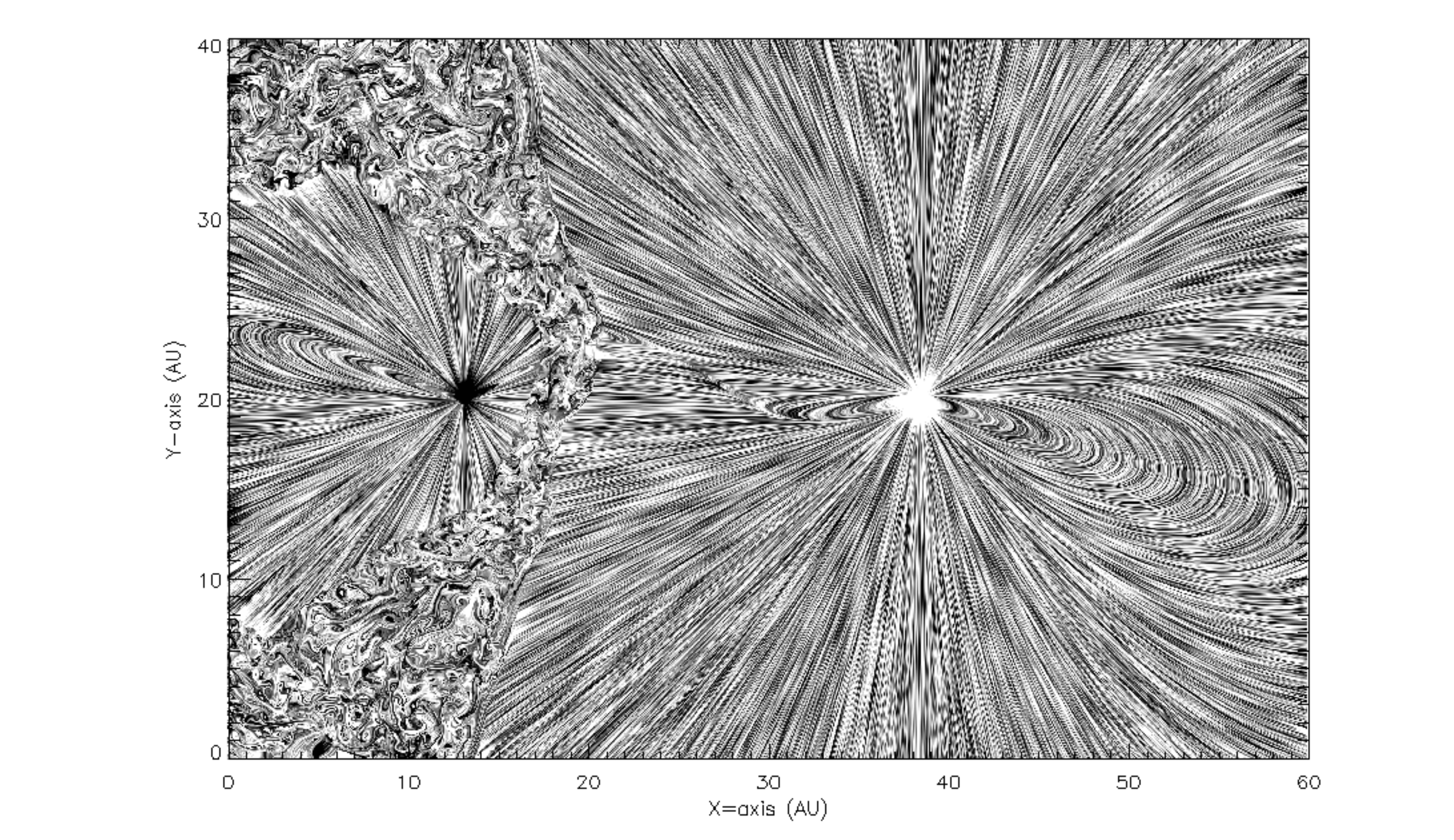}
\caption{Central slices of density (top), magnetic energy density (central) and 
line integral convolution of magnetic field orientation (bottom), for phase 0.5 (apastron).
\label{fig:maps}}
\end{center}
\end{figure}

In Figure \ref{fig:maps} we present the central slice of the model with radiative cooling at 
orbital phase 0.5, i.e. near apastron. The maps are shown for density (top), magnetic energy 
density (central) and line integral convolution (LIC) for the magnetic field orientation (bottom). 
The top panel shows that the standard picture of a two shock layers separated 
by a contact discontinuity is broken. Here the entrainment of material among the two layers is very efficient, 
as a consequence of the Nonlinear Thin Layer Instability. Several observations of the wind-wind collision 
region of massive binaries indicate them to be turbulent \citep{falceta06}. 
This mixing property of turbulent flows is also 
responsible for the complex geometry of the magnetic field lines at the shock. Instead of acumulating at the 
contact discontinuity \citep{silva15}, high magnetic energy density regions are mixed, as shown in the central panel of this figure.

\begin{figure}[H]
\begin{center}
\includegraphics[scale=0.35]{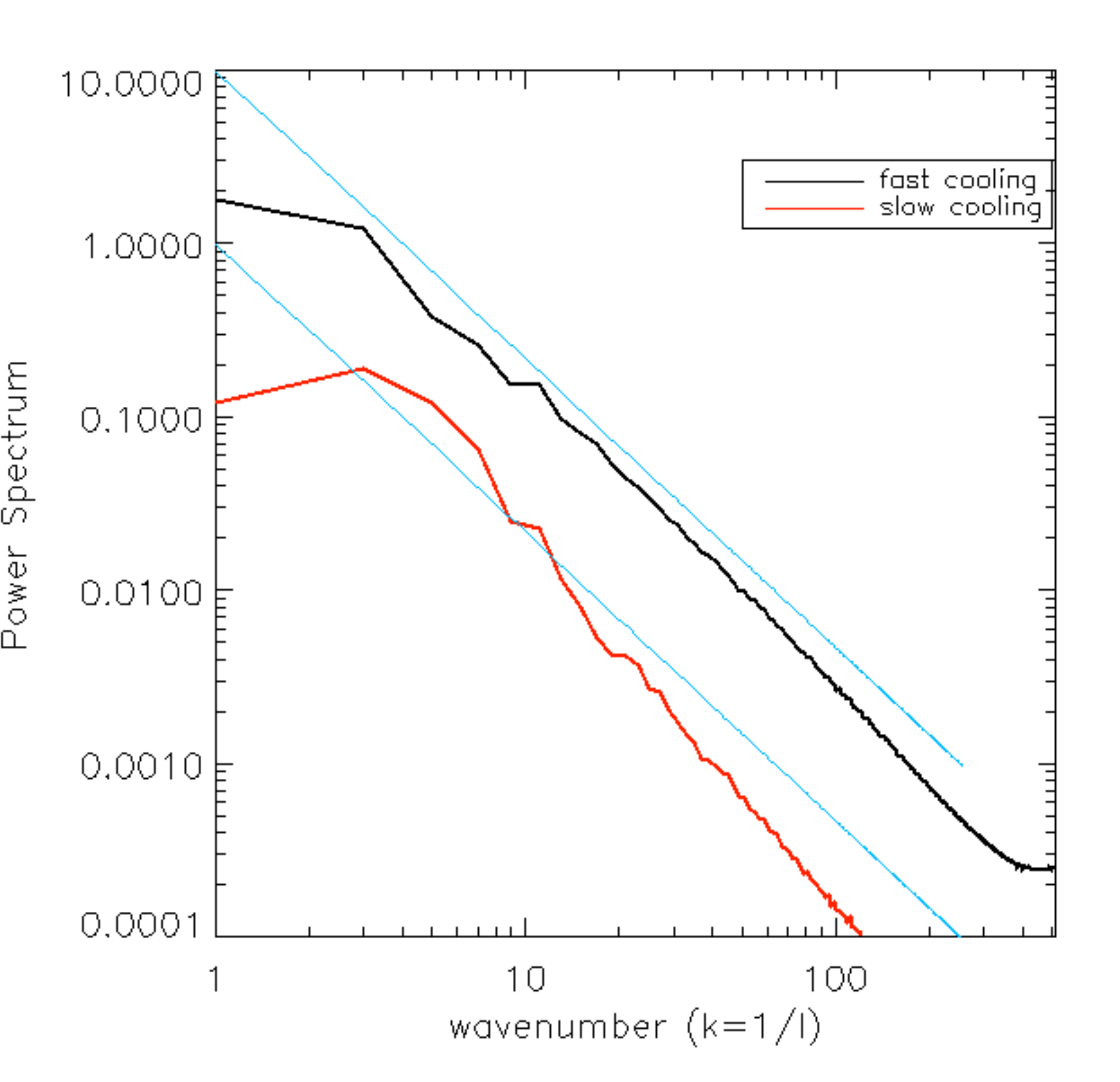}
\caption{Power spectrum of magnetic field for two different models, with 
and without radiative cooling included. Effects of cooling include the 
growth of local instabilities (e.g. nonlinear thin-layer 
instability) and turbulent dynamics.
\label{fig:spectra}}
\end{center}
\end{figure}

The magnetic field orientation at the shock is absolutely chaotic, as seen in the bottom panel of Figure \ref{fig:maps}. Here we show the line integral convolution LIC technique applied for the magnetic field lines. 
The texture shown represents the orientation of the field lines, which are kept approximately dipolar for the 
surroundings of the stars. At the shock, one would expect a field parallel to the shock structure. Indeed, this 
is standard result for the adiabatic model. In the cooled case, turbulence at the shock region results in 
very complex geometry of B-field lines. The scaling of such complex distribution of magnetic field can be obtained 
from its power spectrum.

The turbulent power spectra of the magnetic field at the shock region obtained for the adiabatic 
and cooled models are shown in Figure \ref{fig:spectra}. The plot shows the power spectra obtained 
for the adiabatic and cooled models, and the reference line for Kolmogorov scaling ($P_k \propto k^{-5/3}$). 
It is clear that the cooled model presents much larger power of turbulent components that the adiabatic 
model. Also, the slope obtained for the cooled model resembles well the typical energy cascade of Kolmogorov's turbulence, while the adiabatic model shows a steeper spectrum, as a consequence of a dominance of the more uniform (sheath-like) component \citep{falceta14}.

\subsection{B-n correlation at wind-wind shock regions}

\begin{figure}[H]
\begin{center}
\includegraphics[width=\columnwidth]{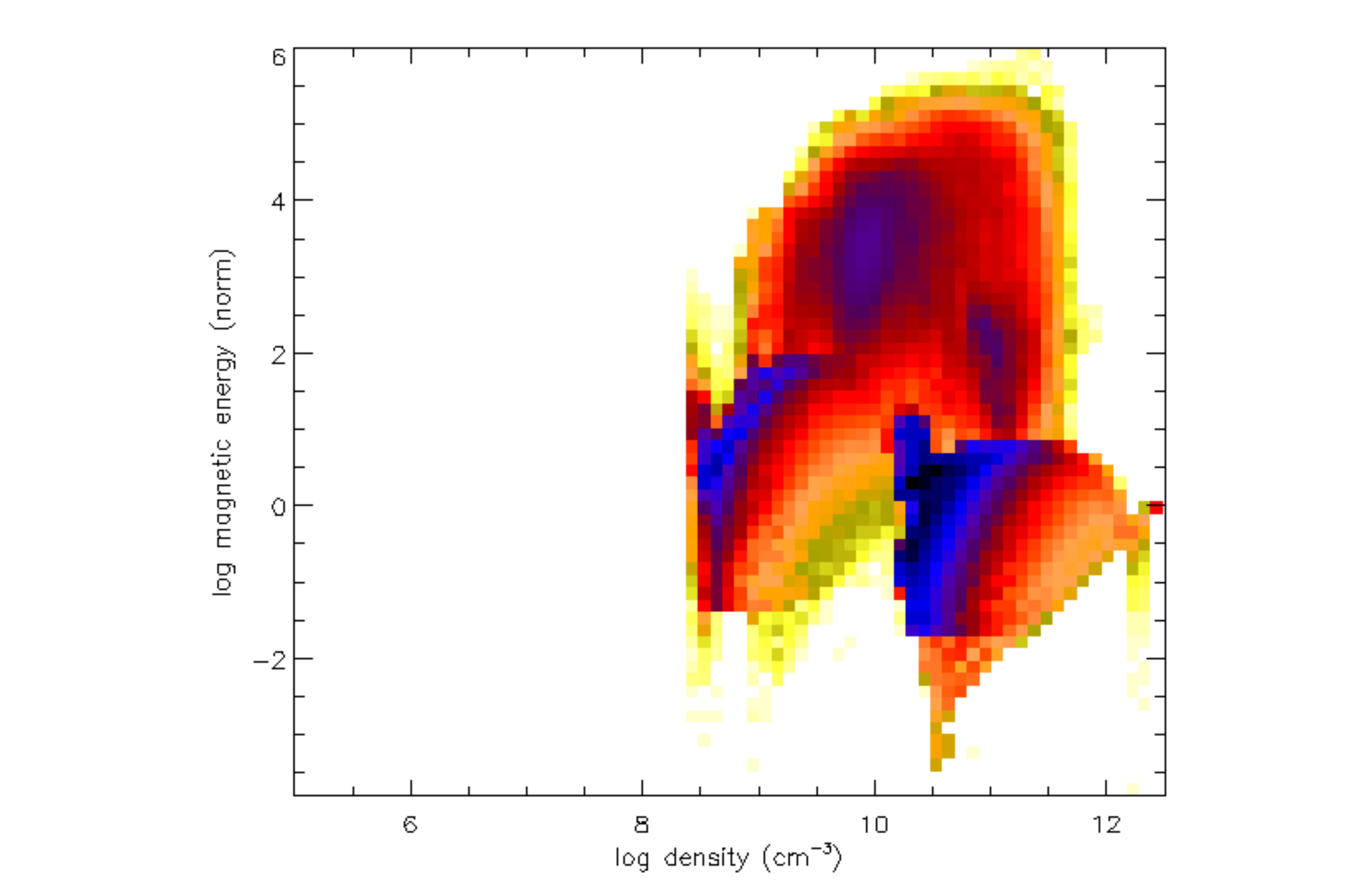}
\caption{2D histogram of the model for the correlation between local density and 
the magnetic energy density. In general, assumption of equipartition $B^2 \propto \rho$ 
is used for the description of the non=thermal component of the shock region (see text), 
however we show that the situation is far more complex.
\label{fig:histogram}}
\end{center}
\end{figure}

The correlation between the magnetic field energy density and the local gas density is 
typically used once equipartition between the magnetic and particle kinectics is assumed. 
This has been largely used by theorists so far in order to model the levels of magnetization 
of shock regions, as well as to predict the high energy content of particle distributions. 
However, as shown in Figure \ref{fig:histogram}, the correlation between $B$ and the gas 
density $n$ is far from trivial. 

Three clustering are visible, representing the stellar winds and the shocked region. The more 
diffuse and highly magnetized cluster of density corresponds to the shock. It is clear that any
attempt to fit a $B \propto n^\alpha$ correlation is unsuccessful \citep{burk09}. Therefore, 
in order to properly model, for instance, the synchrotron emission from massive binaries, one 
needs to fully evolve the magnetic field as well as the acceleration of particles self-consistently.

\subsection{Particle acceleration}

As pointed by \citet{usov93}, the shock region of massive binaries may be 
ideal for particle acceleration to relativistic speeds. 
First order Fermi acceleration is commonly employed in 
such systems, which is efficient once the shock structure is uniform. If after 
each shock crossing the energy of the particle increases by a factor $\delta E/E \sim 2U/3c$, 
where $U$ represents the upstream flow velocity, we obtain - for the parameters of our 
simulations - an acceleration timescale (from keV to TeV) of $\tau_{\rm acc} \sim 10^4-10^5$s. 
Unfortunately, this is slow compared to particle diffusion time ($\tau_{\rm diff} \sim L/v_{\rm part}$), or even for collisions. Therefore, for the model of the system of eta Carinae, the 1st order Fermi process 
is not dominant.

However, we have shown that the shocked region is very turbulent. In such an environment 
particles should be accelerated to relativistic either by the 2nd order Fermi 
process or by electric fields arising from turbulent reconnection zones. In the 2nd order 
process the acceleration occurs as random collisions within the shock region, i.e. diffusion 
at timescales $\tau_{\rm acc} \sim Lc/<\delta v^2>$. For the turbulence parameters obtained 
from the simulations, $<\delta v^2>^{1/2} \sim 100$km s$^{-1}$ at $L \sim 10^9$cm, we obtained 
fast initial acceleration and saturation of relativistic particles with $N(E)dE \propto (E^{-2} - E^{-1})dE$, 
at maximum energy of $1-10$TeV.

\end{document}